# Explicit gauge covariant Euler-Lagrange equation

Clinton L. Lewis

Electronic mail: clinton_lewis@westvalley.edu

*Division of Science and Mathematics, West Valley College, Saratoga, California 95070*

Abstract

The application of a gauge covariant derivative to the Euler-Lagrange equation yields a shortcut to the equations of motion for a field subject to an external force. The gauge covariant derivative includes an external force as an intrinsic part of the derivative and hence simplifies Lagrangians containing tensor and gauge covariant fields. The gauge covariant derivative used in the covariant Euler-Lagrange equation is presented as an extension of the coordinate covariant derivative used in tensor analysis. Several examples provide useful demonstrations of the covariant derivative relevant to studies in general relativity and gauge theory.

## I.    Introduction

The venerable Euler-Lagrange equation continues to provide insights into contemporary physics. The Euler-Lagrange equation extracts the equations of motion for a field from a single function, the Lagrangian. Lagrangian mechanics has the marvelous ability to connect the equations of motion to conservation of momentum, energy, and charge. Examples of equations of motion are Maxwell's equations for electromagnetics, the Klein Gordon equation, the Dirac equation, and other wave equations in space-time. This paper focuses on the equation of motion and how the traditional Euler-Lagrange equation is modified to include a modern generalized derivative. The generalized derivative, the gauge covariant derivative, simplifies the calculations but yields results identical to the traditional Euler-Lagrange equation.

The gauge covariant derivative applies to tensor fields and for any field subject to a gauge transformation. The equations of motion become simpler to derive using the covariant Euler-Lagrange equation because the external gauge field is absorbed by the gauge covariant derivative and hence does not require explicit treatment. The equations of motion automatically include the response to external forces. Only two objects enter the Lagrangian, the field and its derivative. Extending the coordinate covariant derivative to include external gauge fields provides a useful example for students learning gauge theory or tensor analysis. Starting with the familiar example of a complex scalar field and then progressing to examples with vector fields and multiple fields improves accessibility to the topic of covariance.

The entire development in this paper takes place in curvilinear coordinates in space-time and hence conforms to the principle of general covariance, central to general relativity. Using tensors throughout the development in this paper assures that an expression can be written in any coordinate system. Adherence to the principle of general covariance provides additional motivation for finding a covariant form of the Euler-Lagrange equation.

We begin by reviewing the connection between the equations of motion, the Euler-Lagrange equation and the action integral. The gauge covariant derivative is central to this paper and is defined in Sec. IV. The next section is an application of the gauge covariant derivative to a complex field. Section VI derives the central object of this paper, the gauge covariant Euler-Lagrange equation. Further examples in subsequent sections show that the covariant Euler-Lagrange equation remains unchanged for more complicated fields such as vector and for multiple fields. Essentially the same derivation of the covariant Euler-Lagrange equation is presented for three different fields, the repetition emphasizing the differences due to different transformation properties of the fields.

## II.  Background

The essential objects are a field, a derivative of the field, and an external gauge field embedded in the gauge covariant derivative. These objects are functions of space-time coordinates. The field and the derivative of the field are both gauge covariant, meaning that they transform the same way under a gauge transformation. The form of the gauge covariant derivative depends on the gauge transformation properties of the field. The dynamics of the field is defined by an equation of motion. We will see that the equation of motion is extracted from a Lagrangian by means of the Euler-Lagrange equation.

In general relativity partial derivatives with respect to coordinates are replaced by covariant derivatives to arrive at a covariant expression that satisfies the principle of general covariance.[1] Similarly, in the following derivation, partial derivatives are replaced by gauge covariant derivatives to arrive at the covariant Euler-Lagrange equation.[2]

The semicolon is traditionally used to indicate the coordinate covariant derivative of tensor analysis. In this paper it is defined to include both coordinate and gauge covariance properties. The summation convention will be used so that any double index implies summation over space-time coordinates.

## III.  Euler-Lagrange equation from the action integral

Many physical systems are specified by a single function $L$, the Lagrangian. The Euler-Lagrange equation results from applying the principle of least action to the action integral.[3] For example, the action integral of the scalar Lagrangian $L$ over a four-dimensional region $R$ for a system consisting of a complex scalar field $\phi$ acted on by an external electromagnetic field $A_\mu$ is,

$$S = \int_R L(\phi, \partial_\nu \phi, A_\nu) \sqrt{-g} \, d^4 x. \tag{1}$$

The action integral is minimized, or at least stationary, when the scalar field $\phi$ satisfies the Euler-Lagrange equation. The equation of motion of the field is found by applying the Euler-Lagrange equation to a specific Lagrangian. The general volume element in curvilinear coordinates is $\sqrt{-g}\,d^4x$, where $g$ is the determinate of the curvilinear metric.

The electromagnetic vector field $A_\nu$ (a gauge field) is not varied and so is an external field appearing explicitly in the Lagrangian $L(\phi, \partial_\nu \phi, A_\nu)$. The dynamical field $\phi$ has an equation of motion, but the external gauge field does not. After minimization, the Euler-Lagrange equation in its traditional non-covariant form becomes

$$\frac{\partial \sqrt{-g}\,L}{\partial \phi} - \partial_\mu \left( \frac{\partial \sqrt{-g}\,L}{\partial (\partial_\mu \phi)} \right) = 0. \tag{2}$$

This form of the Euler-Lagrange equation provides the equation of motion for the dynamical field specified by the Lagrangian. The partial derivatives of the field which appear in the Euler-Lagrange equation are not gauge covariant, but the resulting equation of motion is gauge covariant as suggested by considering an example such as the Lagrangian for the Klein-Gordon equation.[4]

The simple substitution of the gauge covariant derivative into the Euler-Lagrange equation for the field $\phi$ results in the analogous equation,

$$\frac{\partial L}{\partial \phi} - \left( \frac{\partial L}{\partial \phi_{;\mu}} \right)_{;\mu} = 0. \tag{3}$$

As is demonstrated in the remainder of the paper, Eq. (3) using covariant derivatives is identical to the traditional Euler-Lagrange equation in Eq. (2) using ordinary partial derivatives.

### IV.  *Definition of the gauge covariant derivative*

The gauge covariant derivative is applied to any field responding to a gauge transformation. The essential property of the field is how it transforms, because this property determines the form of the gauge covariant derivative. A few rules help distinguish the gauge covariant derivative from the ordinary partial derivative with respect to a coordinate. In this paper the gauge covariant derivative is indicated by a semicolon.
1) The gauge covariant derivative operates on objects that respond to a coordinate transformation as well as a gauge transformation.
2) The gauge covariant derivative of a field has the same gauge transformation properties as the field itself.
3) A gauge field is embedded in the derivative in order to effect a gauge transformation.
4) The form of the derivative (that is, how the gauge fields are embedded in the derivative) depends on the transformation properties of the field.
5) The derivative satisfies the Leibniz rule (the product rule), $(ab)_{;\mu} = a_{;\mu} b + a b_{;\mu}$.

6) For a gauge invariant scalar, the gauge covariant derivative is identical to the ordinary partial derivative.

The gauge covariant derivative has a specific form which preserves the gauge transformation properties of the field being operated upon, the operand. An additional term containing another field, a gauge field, is required to support the transformation properties. A specific application of the gauge covariant derivative to a complex field follows in the next section.

### V. The covariant derivative applied to the complex field and its conjugate

In our first example we apply the gauge covariant derivative to a complex scalar field $\phi$.[5] The complex scalar field is subject to an external electromagnetic gauge field and could be a wave function that satisfies a relativistic wave equation as the equation of motion. Most importantly, the field and equation of motion for the field transform under a gauge transformation. The gauge covariant derivative is

$$\phi_{;\mu} = \partial_\mu \phi - ieA_\mu \phi, \tag{4}$$

where the electromagnetic external gauge field is $A_\mu$ and the electric charge is $e$. This form of the derivative preserves the same gauge transformation rule (the "covariant" rule) for the covariant derivative as well as for the scalar field itself. The linear gauge transformation rule is

$$\phi \rightarrow e^{ie\theta(x)}\phi, \tag{5}$$

$$\phi_{;\mu} \rightarrow e^{ie\theta(x)}\phi_{;\mu}, \tag{6}$$

with the simultaneous transformation of the electromagnetic vector gauge field.

$$A_\mu \rightarrow A_\mu + \partial_\mu \theta(x). \tag{7}$$

The external gauge field $A_\mu$ has the effect of enabling the desired gauge transformation properties for the covariant derivative. The arrow $\rightarrow$ indicates the replacement for every appearance of the symbol on the left by the expression on the right.

The minus sign changes to a plus for the covariant derivative of the conjugate field $\phi^*$,

$$\phi^*_{;\mu} = \partial_\mu \phi^* + ieA_\mu \phi^*, \tag{8}$$

for the gauge transformation rule (the "contravariant" rule)

$$\phi^* \rightarrow e^{-ie\theta(x)}\phi^*, \tag{9}$$

$$\phi^*_{;\mu} \rightarrow e^{-ie\theta(x)}\phi^*_{;\mu}, \tag{10}$$

and the simultaneous transformation for the gauge field in Eq. (7).

The covariant and contravariant labels can be consistently applied to the field $\phi$ and its complex conjugate $\phi^*$ as is done in tensor analysis. In tensor analysis multiplying a covariant field by a contravariant field results in a scalar field which is invariant under a coordinate transformation.[6] Similarly, the multiplication $\phi^*\phi$ is invariant under a gauge transformation. After identifying $\phi$ in Eq. (5) as a gauge covariant quantity, it follows that $\phi^*$ in Eq. (9) must be gauge contravariant.

### VI. Derivation of the gauge covariant Euler-Lagrange equation

We start by deriving the gauge covariant Euler-Lagrange equation for the complex scalar field $\phi$ and then consider more complicated fields in subsequent sections. The derivation is identical except for the transformation properties of the specific field. Even the differences are strongly related through the concept of the affine connection, to be discussed later.

The gauge covariant derivative of a complex field defined in Sec. V is to be used in the Lagrangian $L(\phi, \phi_{;\nu})$. The gauge field $A_\mu$ appears only in the gauge covariant derivative as indicated in Eq. (4), so that the gauge field is implicit in this form of the Lagrangian.

The following relations establish a conversion between the partial derivatives of the two forms of the Lagrangian without any reference to the action principle. On the right-hand side of the following equations are derivatives of $L(\phi, \phi_{;\nu})$; on the left-hand side are derivatives of $L(\phi, \partial_\nu \phi, A_\nu)$ where the gauge field is explicit. The procedure is to evaluate the ordinary partial derivatives of $L(\phi, \partial_\nu \phi, A_\nu)$ in Eq. (2) in terms of covariant derivatives of $L(\phi, \phi_{;\nu})$ in Eq. (3), thus converting the Euler-Lagrange equation into an equal and covariant form.

We first calculate the partial derivative with respect to the differentiated field using the chain rule.

$$\frac{\partial L(\phi, \partial_\lambda \phi, A_\lambda)}{\partial \partial_\mu \phi} = \frac{\partial L(\phi, \phi_{;\lambda})}{\partial \phi_{;\kappa}} \frac{\partial \phi_{;\kappa}}{\partial \partial_\mu \phi}. \tag{11}$$

The last factor becomes

$$\frac{\partial \phi_{;\kappa}}{\partial \partial_\mu \phi} = \frac{\partial (\partial_\kappa \phi - ieA_\kappa \phi)}{\partial \partial_\mu \phi} = \delta_\kappa^\mu. \tag{12}$$

The substitution of Eq. (12) into Eq. (11) gives

$$\frac{\partial L(\phi, \partial_\lambda \phi, A_\lambda)}{\partial \partial_\mu \phi} = \frac{\partial L(\phi, \phi_{;\lambda})}{\partial \phi_{;\mu}}, \tag{13}$$

which completes the conversion for the partial derivative with respect to the differentiated field.

The undifferentiated field $\phi$ appears by itself and in the gauge covariant derivative so that the derivative of the Lagrangian with respect to the field becomes

$$\frac{\partial L(\phi,\partial_\lambda\phi,A_\lambda)}{\partial \phi} = \frac{\partial L(\phi,\phi_{;\lambda})}{\partial \phi} + \frac{\partial L(\phi,\phi_{;\lambda})}{\partial \phi_{;\kappa}}\frac{\partial \phi_{;\kappa}}{\partial \phi}. \tag{14}$$

The final factor in the third term is,

$$\frac{\partial \phi_{;\kappa}}{\partial \phi} = \frac{\partial(\partial_\kappa \phi - ieA_\kappa \phi)}{\partial \phi} = -ieA_\kappa. \tag{15}$$

We substitute this factor into Eq. (14) and find

$$\frac{\partial L(\phi,\partial_\lambda\phi,A_\lambda)}{\partial \phi} = \frac{\partial L(\phi,\phi_{;\lambda})}{\partial \phi} - ieA_\kappa \frac{\partial L(\phi,\phi_{;\lambda})}{\partial \phi_{;\kappa}}. \tag{16}$$

We continue the conversion by substituting Eqs. (13) and (16) into the Euler-Lagrange equation, Eq. (2),

$$\frac{\partial \sqrt{-g}L(\phi,\phi_{;\lambda})}{\partial \phi} - ieA_\kappa \frac{\partial \sqrt{-g}L(\phi,\phi_{;\lambda})}{\partial \phi_{;\kappa}} - \partial_\mu\left(\frac{\partial \sqrt{-g}L(\phi,\phi_{;\lambda})}{\partial \phi_{;\mu}}\right) = 0. \tag{17}$$

Next we drop the now understood functional dependence in the Lagrangian. Observe that $\sqrt{-g}$ does not depend on the field or its derivatives and so commutes with the partial derivative with respect to the field. We divide through by $\sqrt{-g}$,

$$\frac{\partial L}{\partial \phi} - ieA_\mu \frac{\partial L}{\partial \phi_{;\mu}} - \frac{1}{\sqrt{-g}}\partial_\mu\left(\sqrt{-g}\frac{\partial L}{\partial \phi_{;\mu}}\right) = 0, \tag{18}$$

and substitute the coordinate covariant derivative identity,[7]

$$\nabla_\mu V^\mu = \frac{1}{\sqrt{-g}}\partial_\mu\left(\sqrt{-g}V^\mu\right), \tag{19}$$

to obtain (the $\nabla_\mu$ here refers to the usual coordinate covariant derivative),

$$\frac{\partial L}{\partial \phi} - ieA_\mu \frac{\partial L}{\partial \phi_{;\mu}} - \nabla_\mu\left(\frac{\partial L}{\partial \phi_{;\mu}}\right) = 0. \tag{20}$$

The selection of the correct form of the gauge covariant derivative to be substituted into Eq. (20) depends on whether the partial derivative $\partial L/\partial \phi_{;\mu}$ transforms as $\phi$ or $\phi^*$, that is, is gauge covariant or contravariant. The Lagrangian is assumed to be gauge invariant so that a partial derivative in the parenthesis with respect to the gauge covariant field will be gauge contravariant. To demonstrate contravariance select an arbitrary gauge invariant quantity such as $\phi^*\phi$. Then take the partial derivative with respect to the covariant field $\phi$,

$$\frac{\partial(\phi^*\phi)}{\partial \phi} = \phi^*, \tag{21}$$

so that the entire partial derivative can be seen as transforming contravariantly as $\phi^*$. Similarly, the partial derivative $\partial L / \partial \phi_{;\mu}$ transforms gauge contravariantly as $\phi^*$ in Eq. (9) so that the appropriate form of the gauge covariant derivative is Eq. (8), which has a positive sign in front of the vector field $A_\mu$. We substitute the gauge covariant derivative into Eq. (20), which replaces the coordinate covariant derivative and the term with the gauge field. The Euler-Lagrange equation becomes

$$\frac{\partial L}{\partial \phi} - \left(\frac{\partial L}{\partial \phi_{;\mu}}\right)_{;\mu} = 0, \qquad (22)$$

where the semicolon refers to the gauge covariant derivative that absorbs the term with the explicit gauge field in Eq. (20). Equation (22) is identical to Eq. (3) as was the goal.

Identical steps lead to an equation of motion for $\phi^*$ which is the same as for $\phi$ except that the form of the gauge covariant derivatives are appropriate to $\phi^*$ rather than $\phi$. We see that the partial derivative can be replaced by the gauge covariant derivative to obtain the equations of motion through the Euler-Lagrange equation. This replacement is much simpler than calculating the equations of motion with an explicit appearance of the electromagnetic field in the Lagrangian. In effect, the external gauge field disappears into the gauge covariant derivative. The Euler-Lagrange equation either with a partial derivative, Eq. (2), or with a covariant derivative, Eq. (22), results in the same covariant equations of motion.

### VII.   Generalization to a vector field

The coordinate transformation properties of the scalar field are trivial because its value is invariant under the transformation, and thus for greater complexity we consider the Euler-Lagrange equation for a vector field simultaneously gauge covariant. We will extend the discussion in Ref. 8 by requiring that the vector field also be a gauge covariant field. We add a covariant coordinate index $\nu$ to the field now indicated by $\phi_\nu$ which adds the coordinate transformation property $x \to \bar{x}$,

$$\phi_\nu \to \frac{\partial x^\kappa}{\partial \bar{x}^\nu} \phi_\kappa. \qquad (23)$$

The field now transforms covariantly under a coordinate transformation and the gauge transformation in Eq. (7). The field $\phi_\nu$ can be viewed as a set of independent fields indexed by $\nu$. The coordinate derivative of the field has an additional term containing the Christoffel symbol (or the affine connection[9]), which enables coordinate covariance just as the gauge field enables gauge covariance.

$$\phi_{\nu;\mu} = \partial_\mu \phi_\nu - ieA_\mu \phi_\nu - \Gamma^\rho_{\nu\mu} \phi_\rho \qquad (24)$$

The Euler-Lagrange equation (2) is repeated for each field indexed by $\nu$ because the action integral is minimized for each independent field.

$$\frac{\partial L}{\partial \phi_{\nu}} - \frac{1}{\sqrt{-g}} \partial_{\mu}\left(\sqrt{-g} \frac{\partial L}{\partial(\partial_{\mu}\phi_{\nu})}\right) = 0. \tag{25}$$

The Lagrangian with the explicit appearance of the external gauge field and the Christoffel symbol is now $L(\phi_{\rho}, \partial_{\lambda}\phi_{\rho}, A_{\lambda}, \Gamma^{\rho}_{\nu\mu})$, and the Lagrangian with the implicit appearance is $L(\phi_{\rho}, \phi_{\rho;\lambda})$. As before, the goal is to convert the Euler-Lagrange equation to a covariant form.

The partial derivative of the differentiated field is (sum over $\kappa$ and $\sigma$ to complete the chain rule)

$$\frac{\partial L(\phi_{\rho}, \partial_{\lambda}\phi_{\rho}, A_{\lambda}, \Gamma^{\rho}_{\nu\mu})}{\partial \partial_{\mu}\phi_{\nu}} = \frac{\partial L(\phi_{\rho}, \phi_{\rho;\lambda})}{\partial \phi_{\sigma;\kappa}} \frac{\partial \phi_{\sigma;\kappa}}{\partial \partial_{\mu}\phi_{\nu}}. \tag{26}$$

We substitute Eq. (24) to obtain (compare to Eq. (13))

$$\frac{\partial L(\phi_{\rho}, \partial_{\lambda}\phi_{\rho}, A_{\lambda}, \Gamma^{\rho}_{\nu\mu})}{\partial \partial_{\mu}\phi_{\nu}} = \frac{\partial L(\phi_{\rho}, \phi_{\rho;\lambda})}{\partial \phi_{\nu;\mu}}, \tag{27}$$

which completes the conversion for the partial derivative of the differentiated field.

The undifferentiated field $\phi_{\rho}$ has an extra term (with the Christoffel symbol) compared to Eq. (16).

$$\frac{\partial L(\phi_{\rho}, \partial_{\lambda}\phi_{\rho}, A_{\lambda}, \Gamma^{\rho}_{\nu\mu})}{\partial \phi_{\nu}} = \frac{\partial L(\phi_{\rho}, \phi_{\rho;\lambda})}{\partial \phi_{\nu}} - ieA_{\kappa} \frac{\partial L(\phi_{\rho}, \phi_{\rho;\lambda})}{\partial \phi_{\nu;\kappa}} - \Gamma^{\nu}_{\sigma\kappa} \frac{\partial L(\phi_{\rho}, \phi_{\rho;\lambda})}{\partial \phi_{\sigma;\kappa}}. \tag{28}$$

We continue the conversion by substituting Eqs. (27) and (28) into the Euler-Lagrange equation, Eq. (25) and substituting the tensor identity[10]

$$\Gamma^{\sigma}_{\mu\sigma} = \frac{1}{\sqrt{-g}} \partial_{\mu}\sqrt{-g}, \tag{29}$$

to obtain (compare to Eq. (20))

$$\frac{\partial L}{\partial \phi_{\nu}} - ieA_{\kappa} \frac{\partial L}{\partial \phi_{\nu;\kappa}} - \Gamma^{\nu}_{\sigma\kappa} \frac{\partial L}{\partial \phi_{\sigma;\kappa}} - \Gamma^{\sigma}_{\mu\sigma}\left(\frac{\partial L}{\partial \phi_{\nu;\mu}}\right) - \partial_{\mu}\left(\frac{\partial L}{\partial \phi_{\nu;\mu}}\right) = 0. \tag{30}$$

The definition of the coordinate covariant derivative of a second rank tensor is[11]

$$\nabla_{\mu}\left(\frac{\partial L}{\partial \phi_{\nu;\mu}}\right) = \partial_{\mu}\left(\frac{\partial L}{\partial \phi_{\nu;\mu}}\right) + \Gamma^{\nu}_{\sigma\kappa} \frac{\partial L}{\partial \phi_{\sigma;\kappa}} + \Gamma^{\sigma}_{\mu\sigma}\left(\frac{\partial L}{\partial \phi_{\nu;\mu}}\right). \tag{31}$$

Again, the gauge covariant derivative absorbs the additional term with the gauge field. Finally, the covariant Euler-Lagrange equation becomes

$$\frac{\partial L}{\partial \phi_{\nu}} - \left(\frac{\partial L}{\partial \phi_{\nu;\mu}}\right)_{;\mu} = 0, \tag{32}$$

We have proceeded from the non-covariant Euler-Lagrange Eq. (25) and converted it to the covariant Euler-Lagrange Eq. (32) as we set out to show. This argument can be extended to tensors of any rank because each additional rank adds another term to Eq. (28) so that the covariant Euler-Lagrange equation is equal to the non-covariant version for a field of any rank tensor.

### VIII. Generalization to a multifield

The covariant Euler-Lagrange equation applies in the presence of a more general gauge field as used in gauge theory.[12] Consider the linear transformation of a column matrix of $N$ independent fields $\boldsymbol{\phi}$ (a multifield). Represent the gauge transformation by the $N \times N$ matrix $\mathbf{T}$ and the linear transformation of the field by $\boldsymbol{\phi}$. Latin or Greek bold letters indicate a matrix. The following equations are matrix equations.

$$\boldsymbol{\phi} \to \mathbf{T}\boldsymbol{\phi} \tag{33}$$

$$\boldsymbol{\phi}_{;\mu} \to \mathbf{T}\boldsymbol{\phi}_{;\mu}. \tag{34}$$

The transformation matrix $\mathbf{T}$ may be an element of the $SU(N)$ group, but the development here does not depend on its group properties. The definition of the gauge covariant derivative of the field is

$$\boldsymbol{\phi}_{;\mu} = \partial_\mu \boldsymbol{\phi} - \mathbf{A}_\mu \boldsymbol{\phi}, \tag{35}$$

and the required gauge transformation property of the gauge field, the $N \times N$ matrix gauge field $\mathbf{A}_\mu$, is

$$\mathbf{A}_\mu \to \mathbf{T}\mathbf{A}_\mu \mathbf{T}^{-1} + (\partial_\mu \mathbf{T})\mathbf{T}^{-1}, \tag{36}$$

as can be confirmed by substitution.

The field $\boldsymbol{\phi}$ has multiple components like a tensor, but is arranged as a column matrix. The Euler-Lagrange equation still has the familiar covariant form

$$\frac{\partial L}{\partial \boldsymbol{\phi}} - \left( \frac{\partial L}{\partial \boldsymbol{\phi}_{;\mu}} \right)_{;\mu} = 0. \tag{37}$$

A partial derivative with respect to a column matrix such as $\partial L / \partial \phi_i$, where the specific field is indicated by $i$, is carried out element by element,. The result for this partial derivative with respect to a column matrix is a row matrix.

The essential step in the derivation of the covariant Euler-Lagrange equation is the replacement of the Christoffel symbol in Eq. (28) by the matrix gauge field $\mathbf{A}_\mu$.

$$\frac{\partial L(\boldsymbol{\phi}, \partial_\nu \boldsymbol{\phi}, \mathbf{A}_\nu)}{\partial \partial_\mu \boldsymbol{\phi}} = \frac{\partial L(\boldsymbol{\phi}, \boldsymbol{\phi}_{;\nu})}{\partial \boldsymbol{\phi}} - \frac{\partial L(\boldsymbol{\phi}, \boldsymbol{\phi}_{;\nu})}{\partial \boldsymbol{\phi}_{;\kappa}} \mathbf{A}_\kappa. \tag{38}$$

The remaining steps are similar to the vector field, except that the Christoffel symbols are replaced by the matrix gauge field, and summation over repeating indices is replaced by matrix multiplication. We have now completed the derivation of the covariant Euler-

Lagrange equation for scalar, vector and multiple fields and now investigate an example of a multifield.

## IX. Example of a multifield: The complex scalar field

The complex scalar field appearing in the first example can be seen as a particular case of the multifield case. The electromagnetic vector potential $A_\mu$ is a special case of the gauge field $\mathbf{A}_\mu$ as can be seen by using matrix notation for the covariant derivatives in Eqs. (4) and (8):

$$\begin{pmatrix} \phi \\ \phi^* \end{pmatrix}_{;\mu} = \partial_\mu \begin{pmatrix} \phi \\ \phi^* \end{pmatrix} - \begin{pmatrix} ieA_\mu & 0 \\ 0 & -ieA_\mu \end{pmatrix} \begin{pmatrix} \phi \\ \phi^* \end{pmatrix}. \tag{39}$$

The electromagnetic gauge field $A_\mu$ enters into the matrix $\mathbf{A}_\mu$.

$$\mathbf{A}_\mu = \begin{pmatrix} ieA_\mu & 0 \\ 0 & -ieA_\mu \end{pmatrix}. \tag{40}$$

The gauge transformation defined by Eqs. (5) and (9) becomes

$$\boldsymbol{\phi} \to \mathbf{T}\boldsymbol{\phi}, \tag{41}$$

where the general linear gauge transformation matrix $\mathbf{T}$ for the electromagnetic gauge transformation becomes

$$\mathbf{T} = \begin{pmatrix} e^{ie\theta(x)} & 0 \\ 0 & e^{-ie\theta(x)} \end{pmatrix}. \tag{42}$$

The Klein Gordon complex scalar field is an example of a multifield. The Lagrangian is a gauge invariant scalar and applies to a relativistic wave function for spin zero particles.

$$L = \phi^*_{;\mu} \phi^{;\mu} - m^2 \phi^* \phi. \tag{43}$$

This example uses our construction for the gauge covariant derivative for the complex scalar field which has two independent components. Applying the Euler-Lagrange equation to the Lagrangian, the equation of motion for the Klein-Gordon field is

$$\phi^{;\mu}_{;\mu} = m^2 \phi \tag{44}$$

which includes the external electromagnetic field potential $A_\mu$ ($\hbar = 1$) in the gauge covariant derivative.[13]

$$\left( \partial_\mu - ieA_\mu \right)\left( \partial^\mu - ieA^\mu \right)\phi = m^2 \phi. \tag{45}$$

## X. The affine connection

An interesting side note is that the matrix $\mathbf{A}_\mu$ is analogous to the Christoffel symbol in tensor analysis. The Christoffel symbol and the matrix gauge field $\mathbf{A}_\mu$ can be regarded as specific instances of an affine connection.[14] As evidence of the underlying concept, note that both instances of affine connection have three indices and similar transformation properties.

According to our previously discussed rules, the covariant derivative maintains linear gauge transformations and follows Leibniz rule. The affine connection is essential to these requirements and actually becomes the external gauge field. The specific appearance of the affine connection, whether indexed or a matrix form, depends upon the form of the field the covariant derivative operates on.

The specific form of the gauge covariant derivative, also called minimum coupling, is recognizable with the general appearance of the matrix equation $\phi_{;\mu} = \partial_\mu \phi - \mathbf{A}_\mu \phi$. The ordinary partial derivative is followed by a gauge field in the form of a three indexed object $\mathbf{A}_\mu$ (a square matrix with a coordinate index) multiplying the operand. The indices may be concealed in a matrix form or be exposed in the form of a Christoffel symbol. The more general concept is that the gauge field behaves like an affine connection and is recognizable through its transformation properties.

## XI. Summary

The gauge covariant derivative can be used in the covariant Euler-Lagrange equation, which results in equations of motion identical to the non-covariant form. Both covariant and non-covariant forms of the Euler-Lagrange equation apply to fields with various transformation properties, with the covariant form easier to evaluate. The covariant form of the Euler-Lagrange equation is a result of applying the principle of covariance, and not an accidental result limited to the complex scalar field. The application of the gauge covariant derivative provides a straightforward demonstration of the principle of general covariance, which is central to general relativity, and local gauge covariance, which is important in gauge theory.

### Acknowledgements

I thank Jacques Rutschmann for many helpful comments and encouragement.